\documentstyle[aps,prl,psfig,multicol]{revtex}
\begin{document}
\newenvironment{tab}[1]
{\begin{tabular}{|#1|}\hline}
{\hline\end{tabular}}

\title {Effect of magnetic field in hybrid nanostructures}

\author{N. Stefanakis}
\address{Universit\"at  T\"ubingen,
T\"ubingen Germany}  
\date{\today}
\maketitle

\begin{abstract}
We examine the effect of the magnetic field on the proximity 
effect in nanostructures 
self consistently using the Bogoliubov-deGennes formalism within
the two dimensional extended Hubbard model.
We calculate the local density of states, and 
the pair amplitude. We study several nanostructures: 
superconductor - two dimensional 
electron gas, superconductor - ferromagnet.
In these structures the 
magnetic field
can be considered as a modulation parameter for the 
proximity effect.
\end{abstract}
%\begin{multicols}{2}
\section{Introduction}

When a magnetic field is applied in a two dimensional electron gas (2DEG)
leads to a 
fractal energy spectrum known as the Hofstadter's butterfly, where 
small changes in magnetic field change completely the spectrum 
\cite{hofstadter}. 

Recently conductance measurements have been performed on a 
superconductor 2DEG structure where an external magnetic field 
is applied perpendicular to the 2DEG \cite{moore}. In this case the 
magnetic field is sufficient large to induce Landau quantization 
in 2DEG, however is smaller than the upper critical field of the 
superconductor. More recently experiments have established the 
connection between the Hofstadter's spectrum observed in the 2DEG 
in the presence of a magnetic field and the quantum Hall effect 
\cite{albrecht}.

On the other hand phase sensitive experiments can be used to probe the 
anisotropy of the pair amplitude in
high temperature superconductors
\cite{hilgenkamp,kashiwaya}.
These experiments report the existence of a zero bias
conduction peak (ZBCP) \cite{guillou} which originates from
the zero energy states (ZES) formed near 
the $[110]$ surfaces of 
$d$-wave superconductors \cite{stefan,zhu1}. 

Recently the proximity effect has been probed as
decaying oscillations of the 
density of states in $s$-wave superconductor ferromagnet hybrid 
structures \cite{kontos} and a phase shift of half flux quantum in the 
diffraction pattern of a ferromagnetic $0-\pi$ SQUID \cite{guichard}.
Similar effects have been observed in $d$-wave 
\cite{freamat,freamat1} superconductor ferromagnet hybrid structures.
Theoretical explanation has been given in the framework of the 
quasiclassical theory for $s$-wave \cite{zareyan1} 
and $d$-wave case \cite{zareyan2}. 
In these structures the exchange field
modulates the period of the pair amplitude oscillations. 
Moreover much interest has been focused recently on the 
manipulation of entangled states which are formed by extracting 
Cooper pairs from the superconductor. For example a beam 
splitter has been proposed \cite{lesovik} and also several experiments 
that involve ferromagnetic electrodes connected to 
superconductors \cite{melin,jirari,stefan2}. 
These structures have acquired considerable 
interest the last years due to the possibility to use 
the $\pi$ states in solid state qubit implementation.

In this 
paper 
our goal is to explore several new aspects
related to the control of the proximity effect in
nanostructures. We study superconductor - two dimensional
electron gas, superconductor - ferromagnet.
The basic quantities which we calculate are
the local density of states  and the pair amplitude
as a function of 
several relevant parameters:
the distance from the surface, the magnetic field,
the barrier strength, 
and the symmetry of the 
pair potential. The method is based on exact diagonalizations
of the Bogoliubov-de Gennes equations associated to the mean field
solution 
of an extended Hubbard model.

Our principal result is that for superconductor 2DEG interface 
the LDOS shows a composite picture of energy bands due to 
Landau quantization and gaps due to the presence of 
superconductivity. 
In superconductor ferromagnet hybrid structures the magnetic field
can modulate the period of the pair amplitude oscillations 
inside the ferromagnetic layer.
Our predictions from the simulations of this model are of interest in
view of future STM spectroscopy experiments on nanostructures.

The article is organized as follows. In Sec. II we
develop the model and discuss the formalism. In Sec. III  we discuss
the effect of the magnetic field 
in pure 2DEG, superconductors. 
In Sec. IV  we discuss
the effect of the magnetic field 
and the effect of the strength of the barrier,
in superconductor 2DEG heterostructures. 
In Sec. V we discuss the effect of the magnetic field in a 
superconductor ferromagnet hybrid structure. 
Finally summary and discussions are presented in the last section.

\section{BdG equations, for the superconductor-semiconductor junction 
within the Hubbard model} 

The Hamiltonian of
the extended Hubbard model on a two dimensional square
lattice takes the form
\begin{eqnarray}
H & = & -\sum_{<i,j>}t_{ij}c_{i\sigma}^{\dagger}c_{j\sigma} 
+\mu \sum_{i\sigma} n_{i\sigma}+\sum_{i\sigma} \mu_i^In_{i\sigma} \nonumber \\ 
	  & + & V_0\sum_{i} n_{i\uparrow} n_{i\downarrow}
+\frac{V_1}{2}\sum_{<ij>\sigma\sigma^{'}} n_{i\sigma} n_{j\sigma^{'}}\, 
,~~~\label{bdgH}
\end{eqnarray}
where $i,j$ are sites indices and the angle brackets indicate that the 
hopping is only to nearest neighbors. 
In the presence of a magnetic field the hopping integral $t_{ij}$ 
is modified by the Peierls substitution, which account for the coupling 
of electrons to the magnetic field 
\begin{equation}
t_{ij}(A)=t_{ij}\exp(\frac{i e}{\hbar c } \int_{r_j}^{r_i}A dl).
\end{equation}
$A$ is the vector potential which is chosen in a Landau gauge.
$n_{i\sigma}=c_{i\sigma}^{\dagger}c_{i\sigma}$ is the electron number 
operator at site $i$, $\mu$ is the chemical potential.
$V_0$, $V_1$  
are on site and nearest-neighbor interaction strength.
Within the mean field approximation Eq. (\ref{bdgH})
reduces  to 
the Bogoliubov deGennes equations \cite{gennes}:
\begin{equation} 
\left(
\begin{array}{ll}
  \hat{\xi} & \hat{\Delta} \\
  \hat{\Delta}^{\ast} & -\hat{\xi} 
\end{array}
\right)
\left(
\begin{array}{ll}
  u_{n}(r_i) \\
  v_{n}(r_i) 
\end{array}
\right)
=\epsilon_{n}
\left(
\begin{array}{ll}
  u_{n}(r_i) \\
  v_{n}(r_i) 
\end{array}
\right)
,~~~\label{bdgbdg1}
\end{equation}

such that 
\begin{equation}
\hat{\xi}u_{n}(r_i)=-t\sum_{\hat{\delta}} 
u_{n}(r_i+\hat{\delta})+(\mu^I(r_i)+\mu)u_{n}(r_i)
,~~~\label{bdgxi}
\end{equation}

\begin{equation}
\hat{\Delta}u_{n}(r_i)=\Delta_0(r_i)u_{n}(r_i)+\sum_{\hat{\delta}} 
\Delta_{\delta}(r_i)u_{n}(r_i+\hat{\delta}),~~~\label{bdgdelta}
\end{equation}
where the gap functions are defined by

\begin{equation}
\Delta_0(r_i)\equiv 
V_0<c_{\uparrow}(r_i)c_{\downarrow}(r_i)>,~~~\label{bdgdelta0}
\end{equation}

\begin{equation}
\Delta_{\delta}(r_i)\equiv 
V_1<c_{\uparrow}(r_i+\hat{\delta})c_{\downarrow}(r_i)>,~~~\label{bdgdeltadelta}
\end{equation}
and where $\hat{\delta}=\hat{x},-\hat{x},\hat{y},-\hat{y}$. Equation
(\ref{bdgbdg1}) is subject to the self consistency requirements 

\begin{equation}
\Delta_0(r_i) = \frac{V_0(r_i)}{2}F_0(r_i)= 
\frac{V_0(r_i)}{2}\sum_{n} 
u_{n}(r_i)v_{n}^{\ast}(r_i) 
\tanh\left(\frac{\beta
\epsilon_{n}}{2}\right)
,~~~\label{bdgselfD0}
\end{equation}

\begin{eqnarray}
\Delta_{\delta}(r_i) & = & \frac{V_1(r_i+\hat{\delta})}{2}F_{\delta}(r_i)= \nonumber \\
 &  & \frac{V_1(r_i+\hat{\delta})}{2} \sum_{n} 
(u_{n}(r_i)v_{n}^{\ast}(r_i+\hat{\delta}) + \nonumber \\
 &  & u_{n}(r_i+\hat{\delta})v_{n}^{\ast}(r_i) )\tanh\left(\frac{\beta 
\epsilon_{n\gamma_2}}{2}\right)).~~~\label{bdgselfDdelta}
\end{eqnarray}

We solve the above equations self-consistently.
The numerical procedure has been described elsewhere \cite{stefan,stefan2}. 
We then compute the $d$-wave and the extended $s$-wave gap 
functions given by the expressions:
\begin{equation}
\Delta_d(r_i)=\frac{1}{4}[\Delta_{\hat{x}}(r_i)+\Delta_{-\hat{x}}(r_i)
-\Delta_{\hat{y}}(r_i)-\Delta_{-\hat{y}}(r_i)],~~~\label{bdgdeltad}
\end{equation}
\begin{equation}
\Delta_s^{ext}(r_i)=\frac{1}{4}[\Delta_{\hat{x}}(r_i)+\Delta_{-\hat{x}}(r_i)
+\Delta_{\hat{y}}(r_i)+\Delta_{-\hat{y}}(r_i)].~~~\label{bdgdeltas}
\end{equation}
The pair amplitude for the $s$-wave case is $F_0(r_i)$.
The pair amplitude for the $d$-wave case is given by the 
expression
\begin{equation}
F_d(r_i)=\frac{1}{4}[F_{\hat{x}}(r_i)+F_{-\hat{x}}(r_i)
-F_{\hat{y}}(r_i)-F_{-\hat{y}}(r_i)].~~~\label{bdgFd}
\end{equation}
The local density of states (LDOS) at the $i$th site is given by
\begin{equation}
\rho_i(E)=-2\sum_{n} 
\left [ |u_{n}(r_i)|^2 f^{'}(E-\epsilon_n) 
+ |v_{n}(r_i)|^2 f^{'}(E+\epsilon_n) \right ]
,~~~\label{bdgdos}
\end{equation}
$f^{'}$ is the derivative of the Fermi function,
\begin{equation}
f(\epsilon)=\frac{1}{\exp(\epsilon/k_B T) + 1}
.
\end{equation}

\section{effect of the magnetic field}

We start our investigation of heterostructures
by discussing the effect of the magnetic field 
on the local density of states for a 2DEG.
We consider a two dimensional system of $20\times 20$ sites, and we suppose 
fixed boundary conditions by setting the impurity potential $\mu^{I}=100t$
at the surface.
The temperature is $k_B T=0.1t$.

\begin{figure}
\begin{center} 
\leavevmode 
\psfig{figure=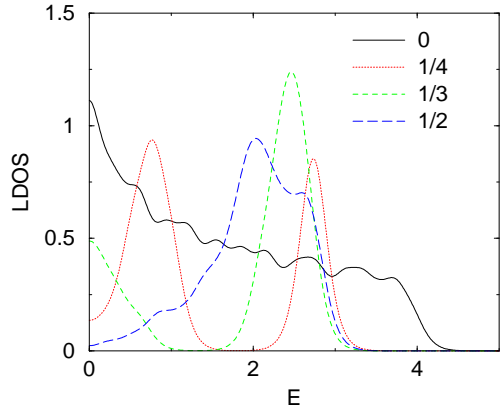,width=6.5cm,angle=0}
\end{center} 
\caption{
The LDOS  
as a function of energy, for a two dimensional electron gas, 
for different values of the magnetic field $f=0,1/4,1/3,1/2$.
}  
\label{2deg.fig}
\end{figure}

The LDOS
as a function of energy, for a two dimensional electron gas,
for different values of the magnetic field is shown in 
Fig. \ref{2deg.fig} (we show only the $E>0$ part of the spectrum 
due to symmetry). We see that when the magnetic flux through 
a plaquette is a rational number $p/q$ the energy spectrum has 
$q$ sub-bands. At each half filled sub-band there is a logarithmic 
singularity. 
The spectrum obtained corresponds to the Hofstadter's butterfly.

\begin{figure}
\begin{center} 
\leavevmode 
\psfig{figure=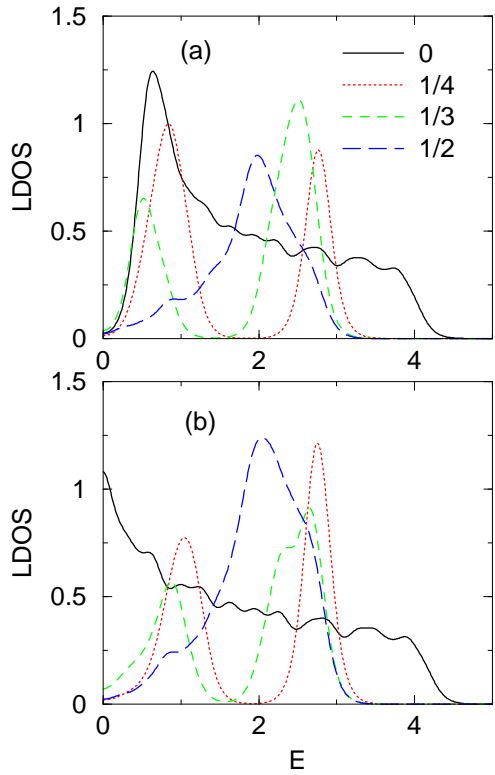,width=6.5cm}
\end{center} 
\caption{
(a) The LDOS  
as a function of energy, for an $s$-wave superconductor, 
for different values of the magnetic field $f=0,1/4,1/3,1/2$.
(b) The same for $d$-wave.
}  
\label{supra.fig}
\end{figure}

We also study the influence of the magnetic field 
in a pure superconducting state.
We assume that the penetration depth is very large so that the 
magnetic field penetrates inside the superconductor.
For the calculation we have chosen 
a Landau gauge for the vector potential, which neglects the effects 
of diamagnetic screening supercurrents. In a more realistic 
calculation the effect of the magnetic field should be calculated 
self-consistently.
The LDOS for an $s$-wave superconductor in the presence of magnetic field 
shows additional gap at energy equal to zero (see Fig. \ref{supra.fig}(a)). 
However the 
above gap LDOS is not modified. 
For the case of $d$-wave superconductor the energy spectrum changes 
for the region inside the gap and also the overall shape of the 
Hofstadter's spectrum is modified qualitatively compared to the 
$s$-wave superconductor (see Fig. \ref{supra.fig}(b)).
 
\section{Hofstadter's spectrum in a superconductor 2DEG heterostructure}

\begin{figure}
\begin{center}
\leavevmode
\psfig{figure=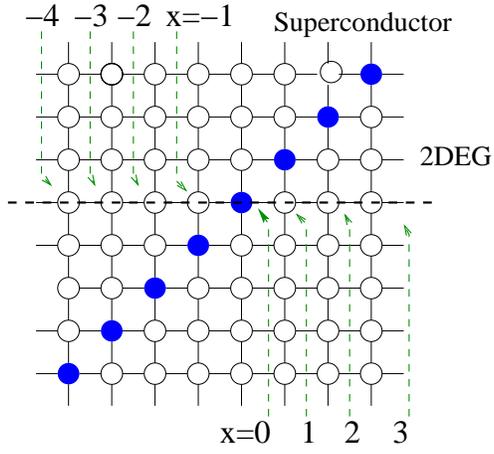,width=6.5cm}
\end{center}
\caption{
The superconductor - 2DEG interface. Solid circles indicate 
interface sites. The LDOS is presented for sites along 
the dashed line. The 2DEG extends to the right of the interface.
} 
\label{interface.fig}
\end{figure}

\begin{figure}
\begin{center}
\leavevmode
\psfig{figure=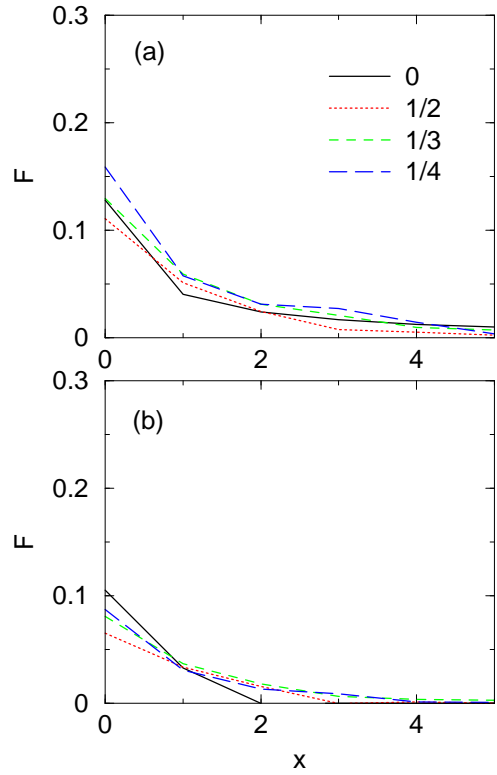,width=6.5cm}
\end{center}
\caption{
(a) The pairing amplitude 
as a function of position, for a $s$-wave superconductor - 
2DEG interface,     
for different values of the magnetic field $f=0,1/4,1/3,1/2$.
(b) The same but for a $d$-wave superconductor.
} 
\label{supra_2degs.fig}
\end{figure}

We now discuss the modification of the Hofstadter's spectrum
in a superconductor 
- 2DEG heterostructure shown in Fig. \ref{interface.fig}
due to the magnetic field. We note that the 
magnetic field is applied only in the 2DEG. When the magnetic field 
is absent, the pair potential 
decays exponentially inside the 2DEG both for $s$-wave and $d$-wave 
superconductor (see Fig. \ref{supra_2degs.fig}). As we increase the exchange 
field the pair amplitude is not really modified inside the 2DEG. 
However inside the superconductor the pair amplitude increases for 
$s$-wave and does not change for $d$-wave.

\begin{figure}
\begin{center}
\leavevmode
\psfig{figure=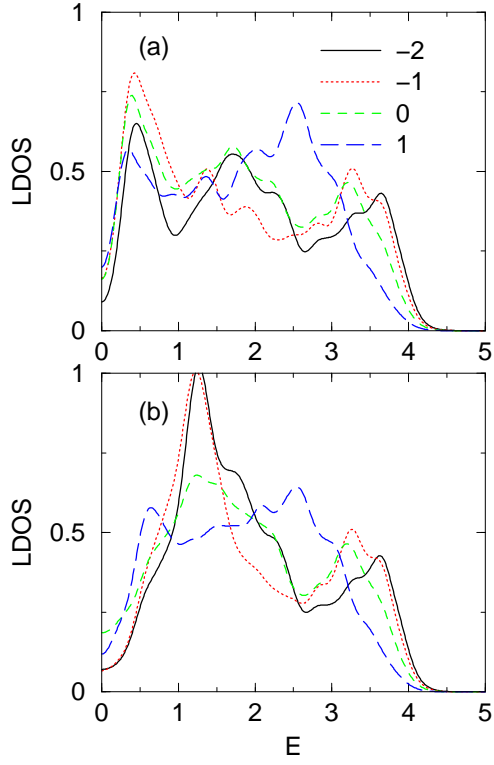,width=6.5cm}
\end{center}
\caption{
(a) The LDOS
as a function of energy, for an $s$-wave superconductor - 
2DEG interface,     
for different values of distance from the interface 
$x=-2,-1,0,1$,
for magnetic field equal to $f=1/2$.
(b) The same but for a $d$-wave superconductor.
} 
\label{supra_2degdos12.fig}
\end{figure}

The proximity effect is even more obvious in the LDOS 
(see Fig. \ref{supra_2degdos12.fig}). We see that for magnetic field 
equal to $f=1/2$, the site inside the superconductor $(-2)$ shows 
a gap structure as a signature of superconductivity 
and also a peak at $E=2$ which is due to the effect of the 
magnetic field. As we move to the 
interior of the 2DEG the gap disappears while the peak the 
at $E=2$ becomes more pronounced. The effect is similar to the 
$s$-wave and the $d$-wave case. So we may conclude that the 
Hofstadter's spectrum can also be observed even in the absence of 
external magnetic field, but only due to the proximity effect with a superconductor, 
and also that the same spectrum is strongly modified in the presence 
of superconducting correlations. 

\begin{figure}
\begin{center}
\leavevmode
\psfig{figure=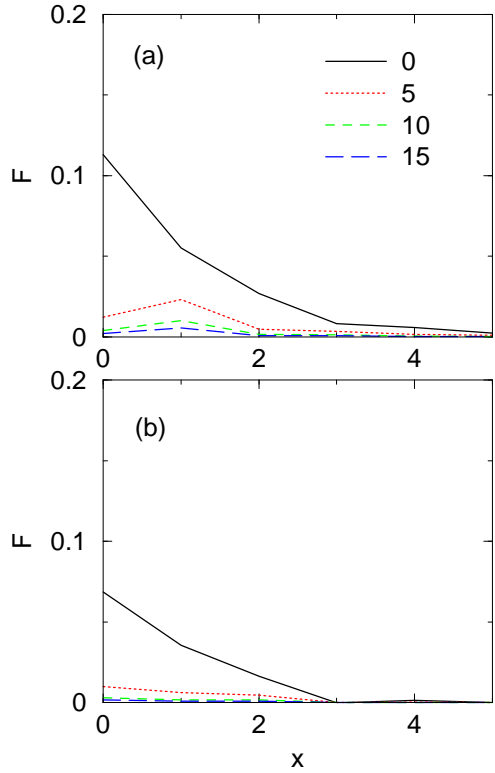,width=6.5cm}
\end{center}
\caption{
(a) The pair amplitude
as a function of distance from the interface, 
for a $s$-wave superconductor - 
2DEG interface,     
for different values of barrier strength
$0,5,10,15$,
for magnetic field equal to $f=1/2$.
(b) The same but for a $d$-wave superconductor.
} 
\label{supra_2degsbar.fig}
\end{figure}

In general the increase of the inter-facial barrier potential
suppresses the 
proximity effect because the leakage of Cooper pairs from the
superconductor to the 2DEG is reduced if the
tunnel amplitude is reduced. In this case the magnetic field 
is screened by the barrier.
As a consequence
for both $s$-wave and 
$d$-wave the pair amplitude is
reduced on the 2DEG side by the increase of the 
barrier strength as seen in Fig. \ref{supra_2degsbar.fig}. 
In the $s$-wave case (see Fig. \ref{supra_2degsbar.fig}(a))
the pair amplitude on the 
superconducting side increases by the increase of the 
barrier strength since the barrier decouples the superconductor 
from the region where the magnetic field is applied,
and the influence of the magnetic 
field in the superconductor is of reduced strength.
In the $d$-wave case on the 
other hand the pair amplitude close to the interface 
does not change much, since the interface becomes pair breaking.

\section{effect of the magnetic field in superconductor ferromagnet 
hybrid structures}

\begin{figure}
\begin{center}
\leavevmode
\psfig{figure=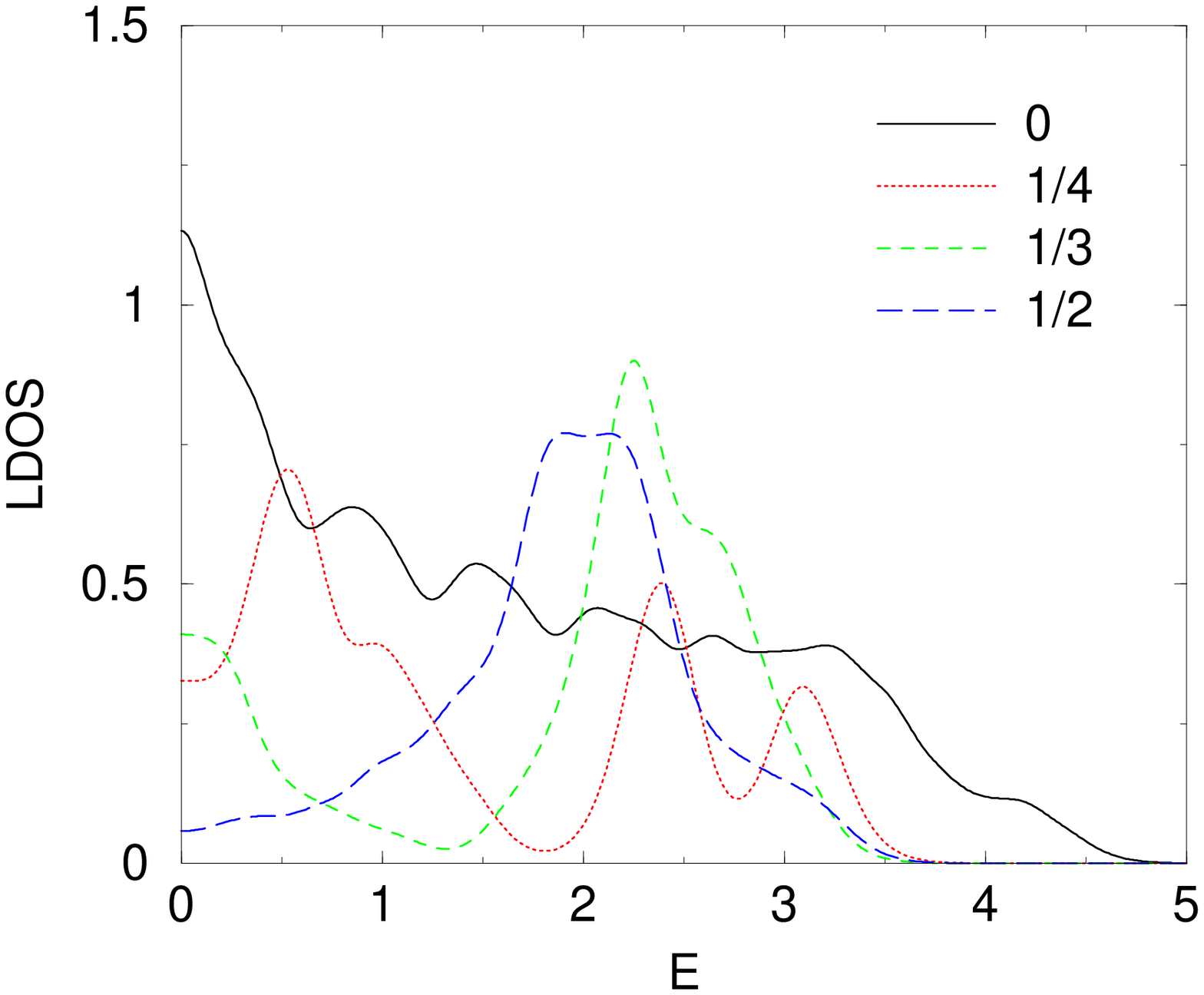,width=6.5cm,angle=0}
\end{center}
\caption{
The LDOS
as a function of energy, for a two dimensional electron gas,
for different values of the magnetic field $f=0,1/4,1/3,1/2$,
for exchange field equal to $h=0.5$.
}
\label{2degh0.5.fig}
\end{figure}

We demonstrate in this section that the magnetic field can act 
as a control parameter for the proximity effect in 
hybrid structures. We study first the effect of the 
magnetic field in a pure ferromagnetic state. 
Due to the presence of the exchange field in the ferromagnet 
the energy bands in the LDOS that are formed due to the 
presence of the magnetic field are spitted (see Fig. \ref{2degh0.5.fig}).

\begin{figure}
\begin{center}
\leavevmode
\psfig{figure=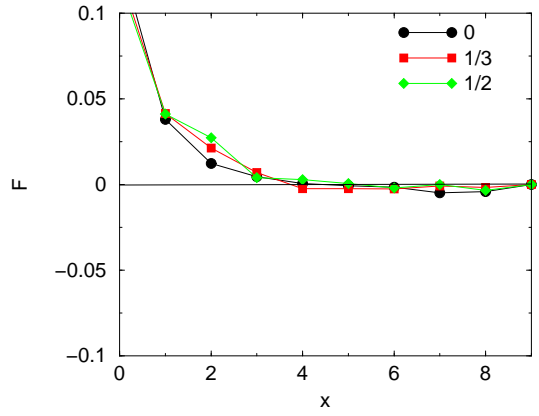,width=6.5cm,angle=0}
\end{center}
\caption{
The pairing amplitude
as a function of position, for a superconductor 
- two dimensional electron gas interface,
for different values of the magnetic field $f=0,1/3,1/2$,
for exchange field equal to $h=0.5$.
}
\label{pah0.5.fig}
\end{figure}

In hybrid structures in the absence of magnetic field, there 
exists oscillations of the pair amplitude in the 
ferromagnetic layer. 
The magnetic field changes the period of the oscillations of the 
pair amplitude as seen in Fig. \ref{pah0.5.fig}. This happens 
because the magnetic field favors single spin orientation 
and therefore modifies the exchange field of the ferromagnet 
to a different effective exchange field. When the magnetic 
field increases the effective exchange field increases and 
the period of oscillations of the pair amplitude decreases 
as seen in Fig. \ref{pah0.5.fig}. 

\begin{figure}
\begin{center}
\leavevmode
\psfig{figure=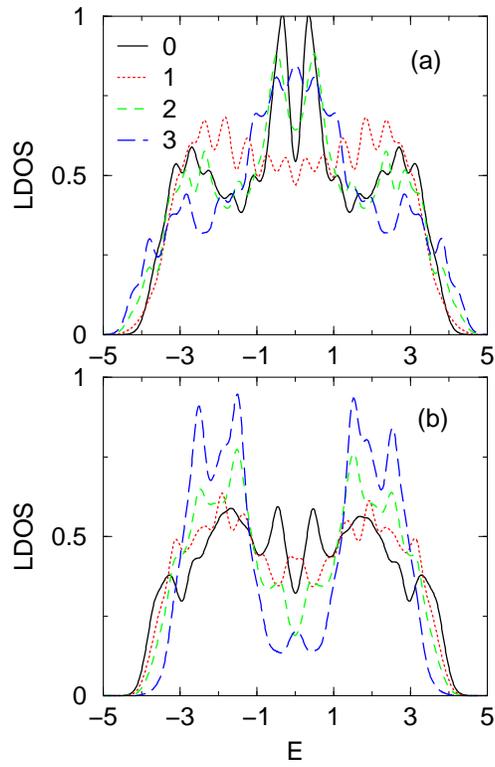,width=6.5cm,angle=0}
\end{center}
\caption{
(a) The LDOS
as a function of energy, for a superconductor - 
two dimensional electron gas interface,
for different values of the distance from the interface $x=0,1,2,3$
for exchange field equal to $h=0.5$, and magnetic field $f=0$.
(b) The same as in (a) but for $f=1/2$.}
\label{ldosh0.5.fig}
\end{figure}

The proximity effect in hybrid structures appears also 
as oscillations in the local density of states with the 
distance from the interface. The external magnetic field 
modifies the period of these oscillations.
We can see the effect of the magnetic field in the LDOS 
in Fig. \ref{ldosh0.5.fig}. We see that for $f=0$ 
the pic in the LDOS which is a signature of the $\pi$ state 
appears after the $x=3$ site from the interface, while 
for $f=1/2$ it appear for smaller distance, because the 
effective period of the oscillations has been changed 
due to the presence of the magnetic field. 

\section{conclusions}
We calculated the LDOS and the pair amplitude for several 
superconductor - 2DEG heterostructures, in the presence of an 
external magnetic field
within the extended Hubbard 
model, self consistently. 
In the calculated quantities, like the pair amplitude and the 
LDOS, for the sites that are close to the interface, the gap which 
is a signature of superconductivity coexists with the 
bands which are formed due to the rational values of the magnetic 
field. We also demonstrated that the Hofstadter's spectrum 
is strongly modified due to the proximity effect with 
the superconductor.
In superconductor ferromagnet structures 
the magnetic field can also be used as an external control 
parameter in order to modify the proximity effect.

\bibliographystyle{prsty}

%\end{multicols}

\end{document}